\documentclass[journal=aelccp,manuscript=letter]{achemso}

\usepackage[version=3]{mhchem}
\usepackage{graphicx}
\usepackage{amsmath,amssymb}
\usepackage[dvipsnames]{xcolor}
\usepackage{siunitx}
\usepackage{multirow}
\usepackage{tabularx}
\usepackage[utf8]{inputenc}
\usepackage[english]{babel}
\usepackage[T1]{fontenc}
\usepackage{caption}
\usepackage{subcaption}
\usepackage{gensymb}
\usepackage{hyperref}
\usepackage{pdfpages}
\graphicspath{{./images/}}

\DeclareUnicodeCharacter{2260}{\neq}
\DeclareUnicodeCharacter{2212}{-}
\DeclareUnicodeCharacter{E5F8}{--}

\author{Syed Mustafa Shah}
\altaffiliation{These authors contributed equally to this work.}
\affiliation[UIC]
{Department of Chemical Engineering, University of Illinois Chicago, Chicago, IL 60608, USA}

\author{Mohammed Lemaalem}
\altaffiliation{These authors contributed equally to this work.}
\affiliation[UIC]
{Department of Chemical Engineering, University of Illinois Chicago, Chicago, IL 60608, USA}
\alsoaffiliation[Argonne]
{Materials Science Division, Argonne National Laboratory, Lemont, IL 60439, USA.}

\author{Anh T. Ngo}
\email{anhngo@uic.edu}
\affiliation[UIC]
{Department of Chemical Engineering, University of Illinois Chicago, Chicago, IL 60608, USA}
\alsoaffiliation[Argonne]
{Materials Science Division, Argonne National Laboratory, Lemont, IL 60439, USA.}

\title{Solvation Restructuring Accelerates Early SEI Nucleation in Lithium Metal Batteries}
\begin{document}

\begin{abstract}
The development of high-energy-density lithium metal batteries is limited by electrolyte instability and the poorly understood onset of solid-electrolyte interphase (SEI) formation. Here, we use AIMD-trained Deep Potential molecular dynamics to link electrolyte solvation structure to early SEI nucleation in \ce{LiTFSI}/DMC electrolytes. Increasing \ce{LiTFSI} concentration drives the electrolyte from solvent-separated ion pairs toward contact ion pairs and aggregates, with \SI{3.5}{M} showing the strongest anion coordination. This anion-rich environment promotes earlier interfacial Li--F/Li--O bond formation, faster consumption of intact \ce{TFSI-} and solvent, and a denser, LiF/Li$_2$O-rich nascent interphase, in contrast to the more organic-laden, phosphorus/fluorine-based interphase formed by a \SI{1}{M} \ce{LiPF6} reference electrolyte. These results show that bulk solvation architecture biases both the timing and chemistry of early SEI formation and suggest solvation control as a design handle for Li-metal electrolytes.
\end{abstract}

\newpage
\textbf{TOC Graphic}\\

\noindent\includegraphics[width=7cm]{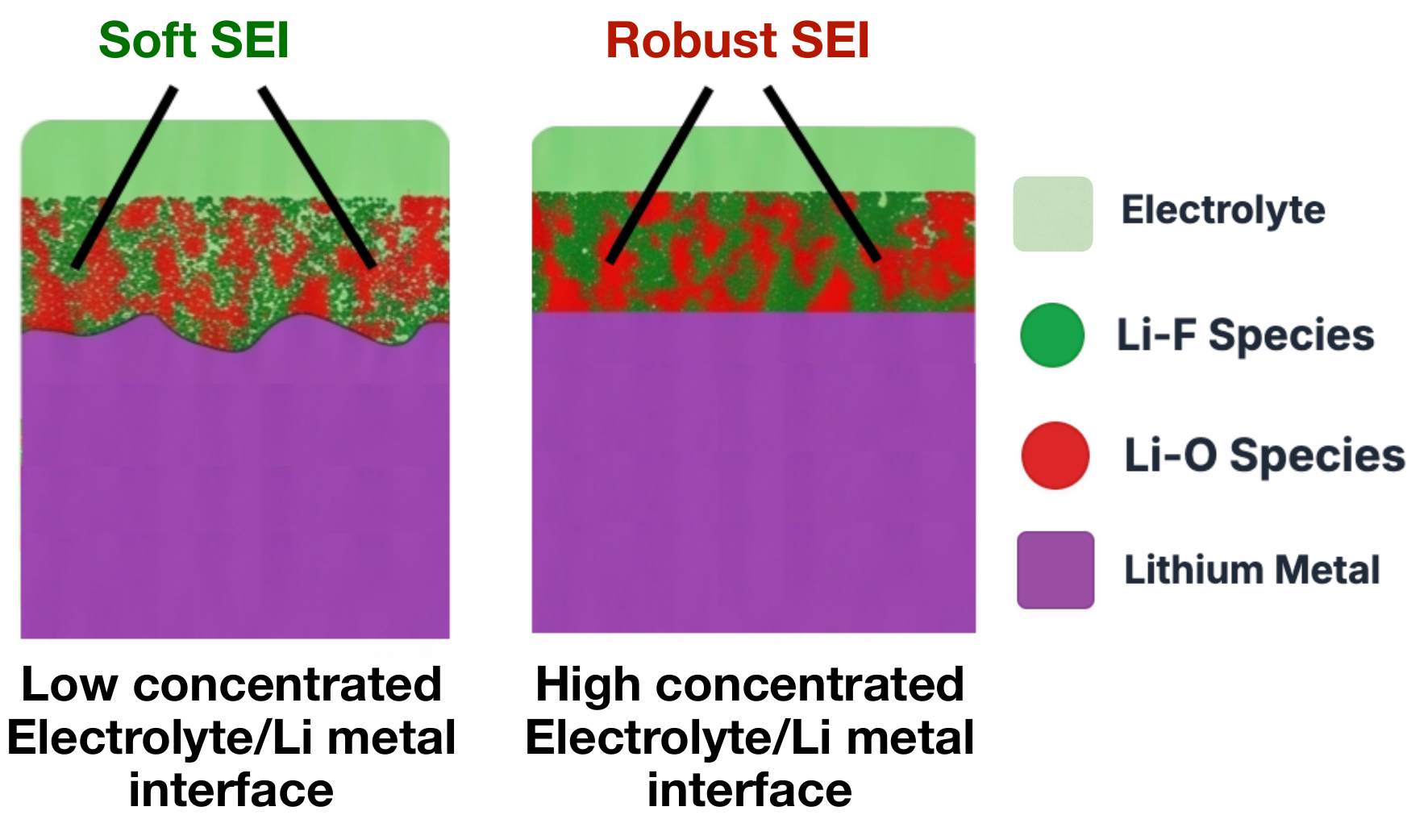}

\section{Introduction}

Lithium metal batteries (LMBs) offer very high energy density but remain limited by electrolyte instability and uncontrolled SEI growth.\cite{lee2026recent,vishweswariah2025evaluation,perez2024sei,Wang2025_RecentTrends} Although recent MLMD studies have enabled larger-scale interfacial simulations \cite{lai2025machine,niu2025large}, the connection between bulk solvation structure and the earliest stages of SEI nucleation remains unclear.\cite{zhou2025tailored,lindsey2026situ,lai2025machine,Zhong2025MLSEI,tao2025deep,fan2024deep,zhang2025recent,wu2025formation}

High-concentration electrolytes (HCEs) are widely used to tune SEI chemistry via solvation engineering.\cite{li2026non,xing2025directing,huang2024crucial} However, connecting bulk solvation to reactive decomposition at the lithium (Li) surface requires simulations that are both chemically accurate and large enough to capture rare early nucleation events. Classical molecular dynamics (MD) captures transport but lacks reactivity,\cite{kuzmina2024structure,jorge2024theoretically,seiferth2023limitations,yao2022applying} whereas \textit{ab initio} molecular dynamics (AIMD) is limited by system size and timescale.\cite{ebadi2016electrolyte, choi2025strategic} Deep Potential-based MLMD bridges this gap, enabling reactive simulations at near-DFT accuracy with system sizes sufficient to probe SEI nucleation in concentrated electrolytes.\cite{zhang2018deep,zeng2023deepmd}

Here, we investigate \SI{1.5}{M}, \SI{2.5}{M}, and \SI{3.5}{M} \ce{LiTFSI}/DMC, with \SI{1}{M} \ce{LiPF6}/EC-EMC-DMC as a qualitative reference. Our DP-based MLMD simulations span \num{20000} atoms and \SI{700}{ps}, probing bulk solvation structure and interfacial reactive dynamics within a single framework. This enables direct connection between concentration-dependent coordination in the electrolyte and sub-100 ps SEI nucleation at the Li surface, identifying dominant decomposition pathways under realistic conditions.

We resolve the critical sub-100 ps nucleation regime, revealing how concentration-dependent solvation alters the balance between anion-driven and solvent-driven decomposition. This work provides a mechanistic basis for electrolyte design linking bulk solvation to interfacial chemistry, beyond prior interface-focused MLMD or bulk-only classical MD studies.

\begin{figure*}[!ht]
  \centering
  \includegraphics[width=0.8\textwidth]{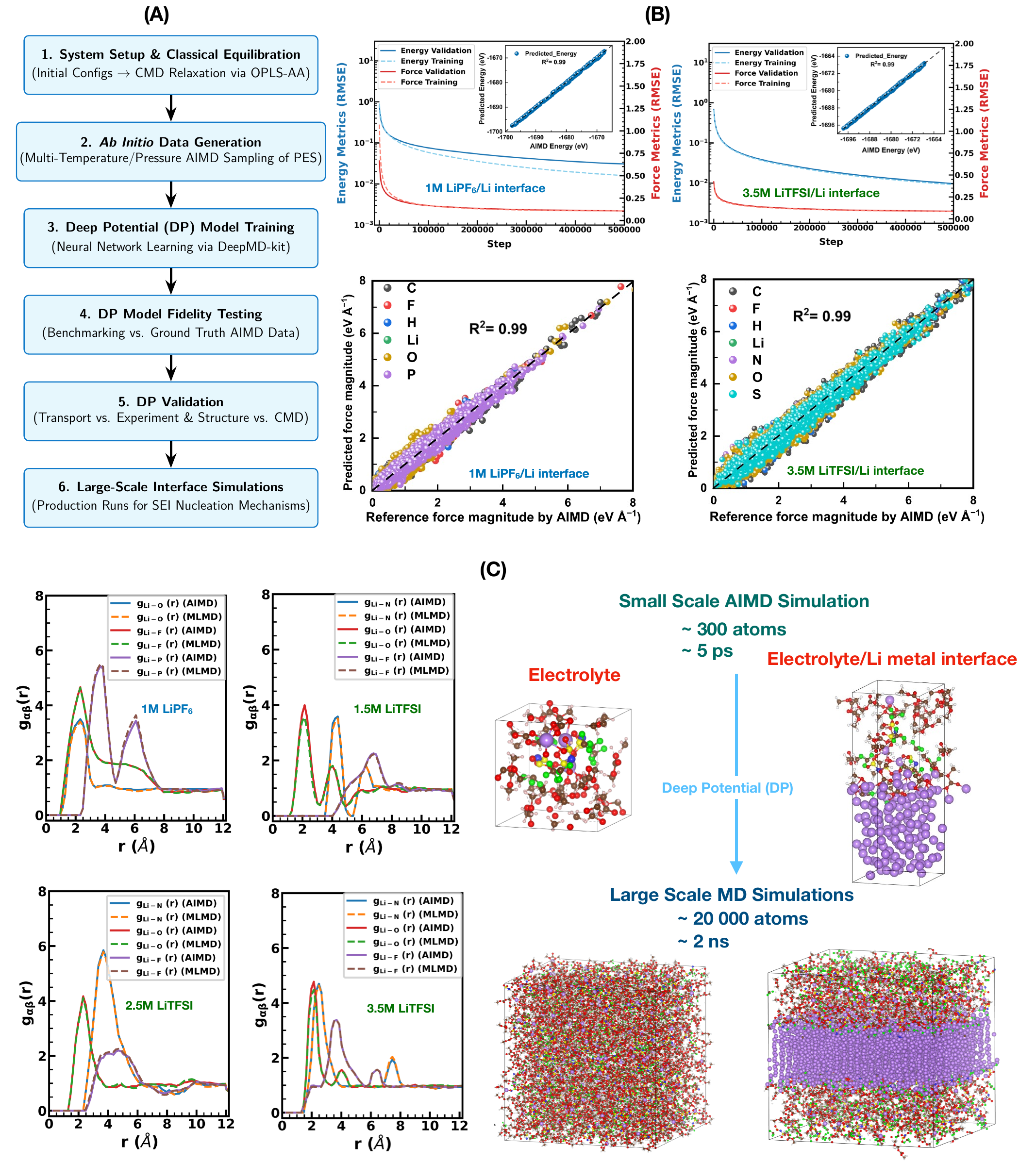}
  \caption{Simulation Framework and Validation. (A) Integrated workflow spanning \textit{ab initio} data generation, Deep Potential (DP) model construction, validation against experimental and classical baselines, and large-scale interfacial simulations. (B) Parity plots comparing DP-predicted energies and forces with AIMD reference data. (C) Radial distribution functions demonstrating structural consistency between MLMD and AIMD trajectories.}
  \label{fig:Fig1}
\end{figure*}

\textbf{Computational Framework.} 
The workflow (Figure~\ref{fig:Fig1}A) combines CMD equilibration, AIMD sampling, DeepMD-kit training, and validation. The training performance (Figure~\ref{fig:Fig1}B) demonstrates rapid convergence of both energy and force RMSE over \(5 \times 10^5\) steps, with minimal gaps between training and validation errors, indicating strong generalization. Parity plots yield \(R^2 \approx 0.99\), and despite the large absolute energy range (approximately \(-1670\) to \(-1700\) eV), the RMSE remains low (\(\sim 10^{-1}\) eV). High accuracy is further confirmed by the strong agreement between AIMD and MLMD RDFs (Figure~\ref{fig:Fig1}C). Methodological details are listed in the Supporting Information.

While the Deep Potential model does not use explicit atomic charges during runtime, charge transfer is encoded implicitly through training on AIMD trajectories at the DFT level. As a result, the model captures thermally activated interfacial reactivity and the chemical instability of these electrolytes against Li metal within the training-informed simulation regime \cite{cheng2025spontaneous,choi2025strategic,beltran2022sei}. The MLMD trajectories further indicate that early-stage SEI nucleation occurs rapidly upon electrolyte--anode contact on the simulated timescale. 

\section{Results and Discussion}

\subsection{Validation of Bulk Solvation Structure and Transport Properties}

\begin{figure*}[!ht]
\centering
\includegraphics[width=0.8\textwidth]{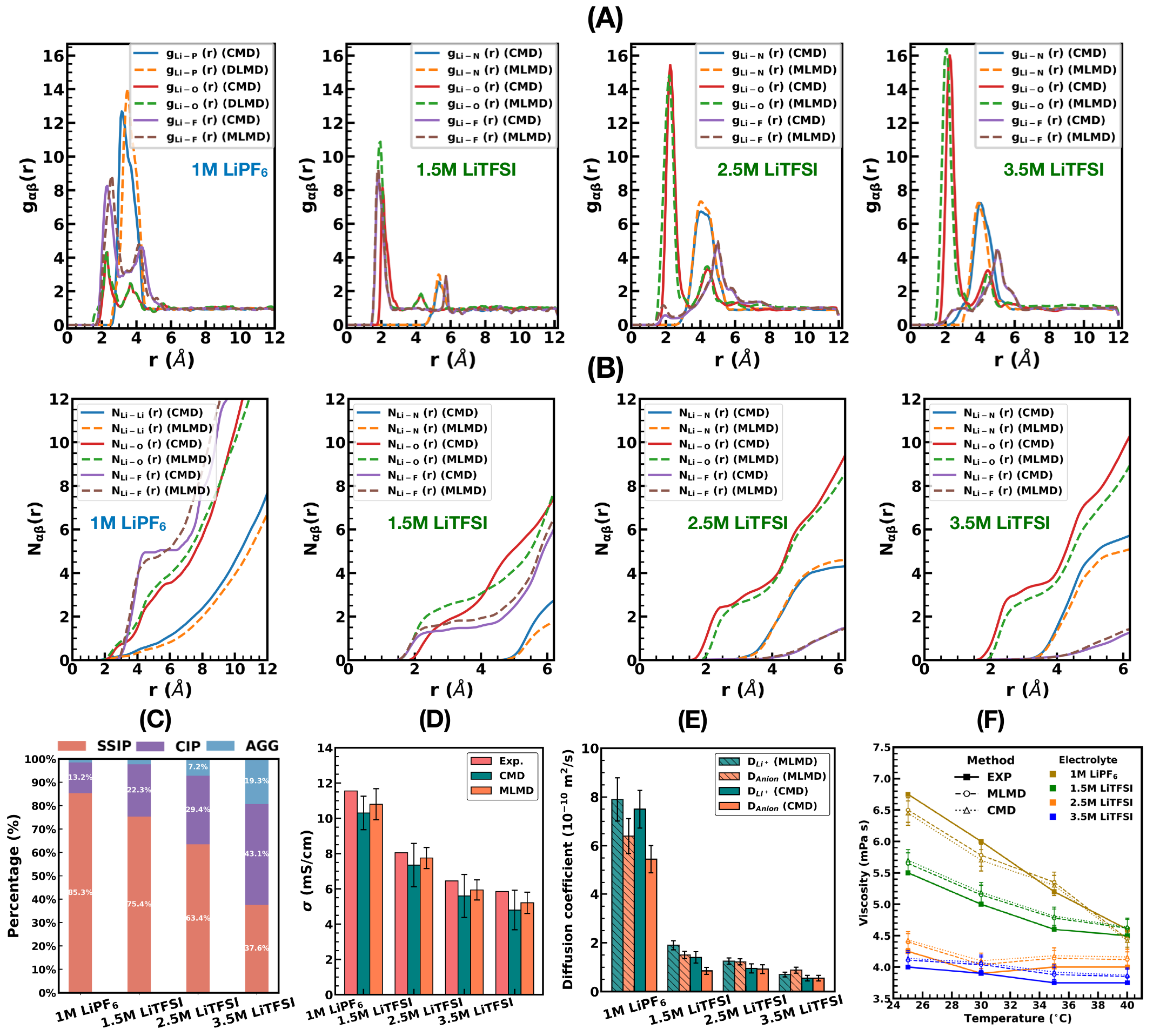}
\caption{Validation of Liquid Structure and Transport Properties. (A) Comparison of radial distribution functions and (B) running coordination numbers predicted by CMD versus MLMD. (C) Distribution of ion-speciation states in different electrolytes, (D) ionic conductivity, (E) ion self-diffusion coefficients, and (F) viscosity. Conductivity and viscosity values are benchmarked against experimental data from Ref.~\cite{li2023concentrated}.}
\label{fig:Fig2}
\end{figure*}

Figure~\ref{fig:Fig2}(A) compares radial distribution functions, \(g_{\alpha\beta}(r)\), for \ce{LiTFSI} and \ce{LiPF6} electrolytes. MLMD and CMD predict similar overall solvation motifs, but differences appear in the first coordination shell. For \ce{LiTFSI}, MLMD predicts shorter Li--O distances than CMD across concentration, indicating a more compact solvation environment. For \ce{LiPF6}, CMD predicts shorter Li--F and Li--P distances than MLMD, while Li--O correlations are reproduced reasonably well by both methods.

The cumulative coordination numbers, \(N_{\alpha\beta}(r)\), in Figure~\ref{fig:Fig2}(B) support these trends. In \ce{LiTFSI}, MLMD shows an earlier rise and larger Li--O coordination than CMD, consistent with a denser primary solvation shell. In \ce{LiPF6}, CMD yields a steeper Li--F coordination curve and a slightly higher plateau. Overall, MLMD captures the concentration-dependent tightening of the \ce{Li+} solvation shell more consistently than CMD for the systems studied here.

The mean squared displacements in Figure~S7 show that \SI{1}{M} \ce{LiPF6} has the highest ion mobility, whereas \ce{LiTFSI}/DMC electrolytes slow progressively as concentration increases to \SI{3.5}{M}, consistent with the reduced ion self-diffusion coefficients reported in Figure~\ref{fig:Fig2}(E). This trend is reflected in the conductivity data in Figure~\ref{fig:Fig2}(D), which decreases with increasing \ce{LiTFSI} concentration and correlates with the higher viscosity in Figure~\ref{fig:Fig2}(F), stronger ion pairing, and reduced \ce{Li+} diffusivity. MLMD agrees better with the experimental conductivity and viscosity trends than CMD, suggesting that the learned potential captures important many-body features relevant to transport in concentrated electrolytes. Although the Deep Potential framework is short-ranged by construction, it provides a useful description of many-body interactions relative to non-polarizable classical force fields. Recent long-range MLIP developments, such as latent-space Ewald descriptors \cite{king2025machine}, offer a path toward further improvement in future work.

\subsection{Relationship Between Solvation Architecture and SEI Nucleation Pathways}

The divergent SEI growth modes observed in the simulations are closely linked to the concentration-dependent evolution of the \ce{Li+} solvation shell. Increasing \ce{LiTFSI} concentration shifts the solvation environment from predominantly solvent-separated ion pairs to contact ion pairs and aggregates, as quantified by the ion-speciation distributions in Figure~\ref{fig:Fig2}(C). In the \SI{1.5}{M} and \SI{2.5}{M} systems, the primary solvation shell is largely occupied by DMC molecules, which separate \ce{TFSI-} from \ce{Li+} and reduce direct anion access to the interface. In contrast, at \SI{3.5}{M} \ce{LiTFSI}, reduced solvent availability promotes closer coordination between \ce{Li+} and the sulfonyl oxygen and fluorine atoms of \ce{TFSI-}. This anion-rich environment is associated with earlier interfacial bond formation and decomposition events when the electrolyte contacts the Li surface, as resolved directly in the interfacial RDF and coordination-number analysis presented below (Figure~\ref{fig:sei_nucleation}).

Because the \ce{LiTFSI} and \ce{LiPF6} systems differ in both salt and solvent composition, the \ce{LiPF6} case is used only as qualitative mechanistic context rather than as a strict control. These trends suggest that concentrated \ce{LiTFSI} favors earlier development of an inorganic-rich nascent interphase, whereas the lower-concentration and cross-salt reference systems more readily form less compact interphases with a relatively larger organic contribution.

\subsection{Atomistic Resolution of Early-Stage SEI Nucleation and Growth}

\begin{figure}[htbp]
    \centering
    \includegraphics[width=\textwidth]{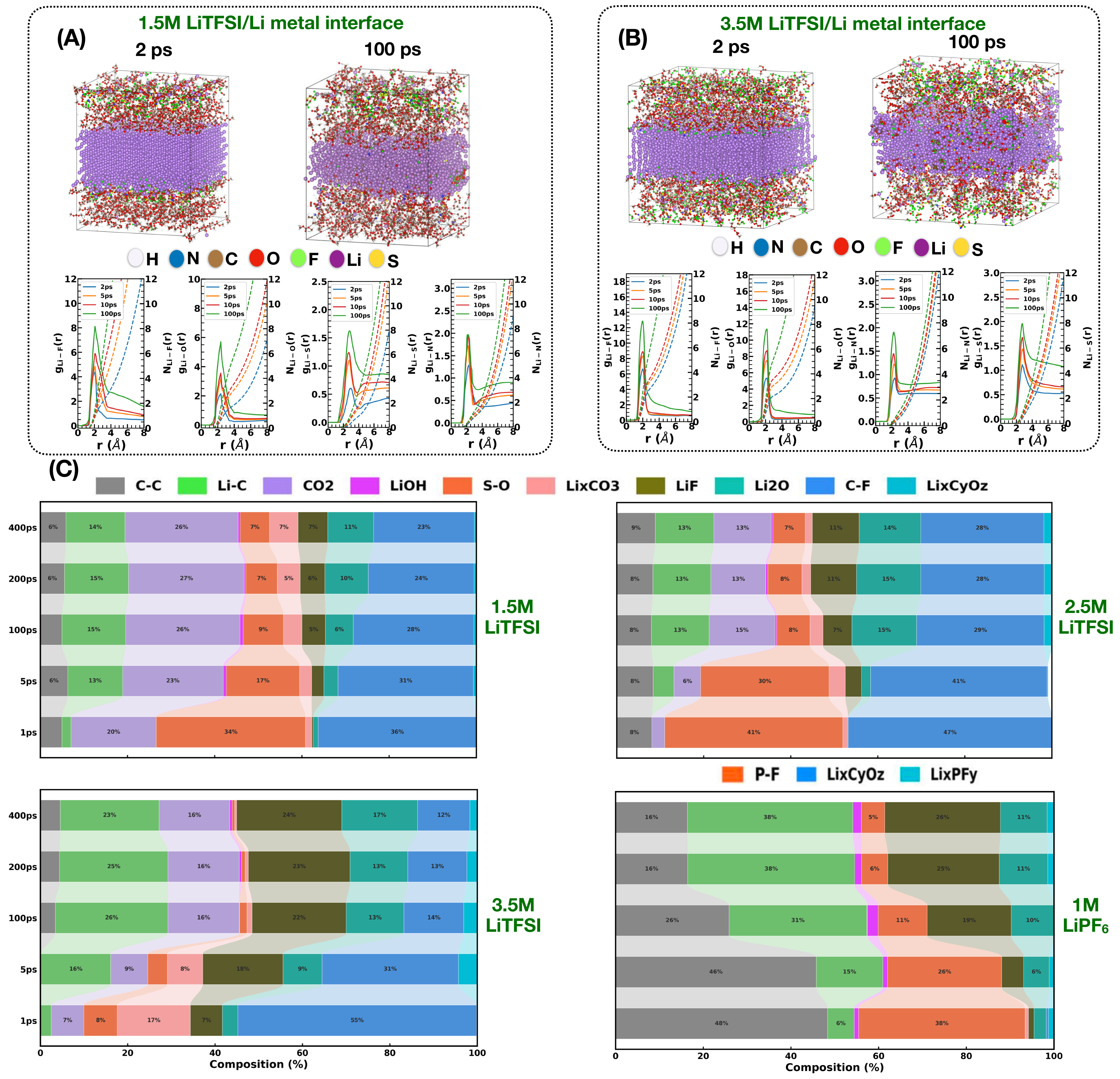}
    \caption{Early-stage SEI nucleation and compositional evolution.
    (A) Representative interfacial snapshots and Li--F, Li--O, Li--N, and
    Li--S radial distribution functions $g(r)$ (dashed lines, left axes)
    with running coordination numbers $N(r)$ (solid lines, right axes) at
    2, 5, 10, and 100~ps for the 1.5~M LiTFSI/Li interface.
    (B) Corresponding snapshots and Li--X RDFs/coordination numbers for
    the 3.5~M LiTFSI/Li interface.
    (C) Temporal evolution of interfacial chemical composition (\%) at
    1, 5, 100, 200, and 400~ps for 1.5~M, 2.5~M, and 3.5~M LiTFSI,
    decomposed into C--C, Li--C, CO$_2$, LiOH, S--O, Li$_x$CO$_3$, LiF,
    Li$_2$O, C--F, and Li$_x$C$_y$O$_z$ motifs, and for 1~M LiPF$_6$,
    decomposed into the analogous C--C, Li--C, and LiOH motifs together
    with the P/F-specific categories P--F, Li$_x$C$_y$O$_z$, and
    Li$_x$PF$_y$, reflecting the distinct decomposition chemistry of the
    \ce{PF6-} anion relative to \ce{TFSI-}.}
    \label{fig:sei_nucleation}
\end{figure}
The compositional evolution in Figure~\ref{fig:sei_nucleation}C provides
a chemical counterpart to these structural trends. At the earliest
times (1--5~ps), all LiTFSI systems are dominated by S--O and C--F
fragments arising from initial reductive attack on the TFSI$^-$ anion.
As simulation time progresses, this transient fragment population
converts toward more stable inorganic products---LiF, Li$_2$O, and
Li$_x$CO$_3$. The rate and extent of this conversion scale with LiTFSI
concentration: by 100--400~ps, the 3.5~M system exhibits the largest
combined LiF/Li$_2$O fraction and the smallest residual S--O
contribution among the LiTFSI series, whereas 1.5~M LiTFSI retains a
larger organic and S--O content over the same window.

Because \ce{PF6-} decomposes through a distinct phosphorus/fluorine
pathway rather than the sulfonyl/fluorine chemistry of \ce{TFSI-}, the
1~M LiPF$_6$ system is tracked using the complementary P--F and
Li$_x$PF$_y$ categories rather than the S--O/LiF/C--F motifs used for
LiTFSI. At 1~ps, the LiPF$_6$ interphase is dominated by C--C-type
solvent fragments (48\%) alongside a substantial P--F contribution
(38\%), reflecting rapid initial anion attack at the Li surface. By
100--400~ps, the composition shifts toward a Li--C-rich interphase
(31--38\%) with persistent P--F (19--26\%) and Li$_x$PF$_y$ (6--11\%)
content, indicating that phosphorus/fluorine-derived species remain a
substantial but not dominant fraction of the nascent interphase.
Relative to the concentrated LiTFSI systems, this composition reflects
a more organic-laden, less densely inorganic nascent interphase, even
though the specific inorganic motif (P--F/Li$_x$PF$_y$) differs
chemically from the LiF/Li$_2$O products that dominate concentrated
LiTFSI.

Together, the structural (RDF/coordination) and compositional trends in
Figure~\ref{fig:sei_nucleation} show that increasing LiTFSI
concentration accelerates both Li$^+$--anion coordination and the
conversion of early decomposition fragments into an inorganic,
LiF/Li$_2$O-rich product distribution. This concentration-dependent
coupling between solvation structure and interfacial chemistry is the
structural basis for the divergent ``soft'' versus ``robust'' SEI
growth modes summarized schematically in Figure~\ref{fig:Fig5}. Overall, increasing \ce{LiTFSI} concentration drives a more rapid transition toward a denser, inorganic-rich nascent interphase, whereas the \SI{1}{M} \ce{LiPF6} system studied here evolves toward a more organic-laden interphase with a comparatively smaller, phosphorus/fluorine-based inorganic contribution. Taken together, these trends indicate that both salt identity and concentration influence interfacial reactivity, with concentration appearing to play the dominant role within the \ce{LiTFSI} series \cite{yang2021robust,alvarado2019bisalt,shadike2022engineering,kuai2023inorganic,gao2025lif}.

\subsection{Interfacial Reduction and Evolution of SEI Architecture}

\begin{figure*}[!ht]
    \centering
    \includegraphics[width=0.8\textwidth]{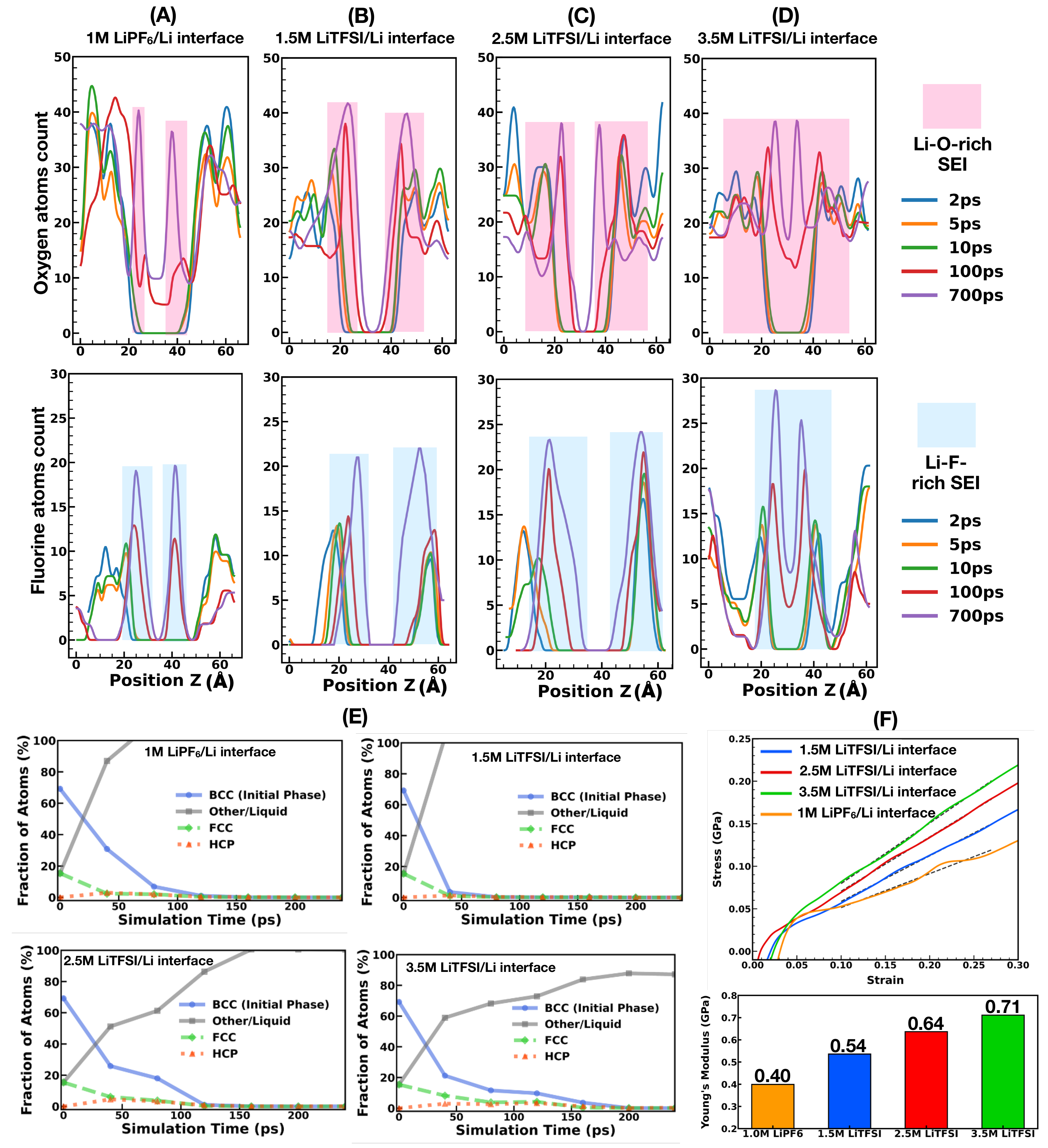}
    \caption{Temporal Evolution of Interfacial Composition. Oxygen and fluorine atomic density profiles along the surface normal ($z$-direction) for (A) \SI{1}{M} \ce{LiPF6}/Li interface, (B) \SI{1.5}{M} \ce{LiTFSI}/Li interface, (C) \SI{2.5}{M} \ce{LiTFSI}/Li interface, and (D) \SI{3.5}{M} \ce{LiTFSI}/Li interface. (E) Evolution of the lithium crystalline structure during early-stage SEI nucleation. (F) Stress-strain curves of the formed SEIs.}
    \label{fig:Fig4}
\end{figure*}

Figure~\ref{fig:Fig4} shows the reorganization of oxygen- and fluorine-containing species along the surface normal during early interphase nucleation and growth. The oxygen density profiles in Figure~\ref{fig:Fig4}(A--D) exhibit a two-stage evolution. At early times (\SIrange{2}{5}{ps}), sharp interfacial peaks appear, indicating rapid accumulation of solvent- and anion-derived fragments at the boundary. By \SI{100}{ps}, these peaks broaden and shift toward the Li-metal region, suggesting that oxygen-rich SEI products extend into the near-surface region. This behavior is most pronounced for \SI{3.5}{M} \ce{LiTFSI}, whereas \SI{1}{M} \ce{LiPF6} shows a more diffuse oxygen distribution and a less compact interphase.

The fluorine density profiles follow a similar trend. Initial narrow peaks broaden and, in some cases, split over time, consistent with the transition from adsorbed intact anions to a fluorine-enriched interphase containing LiF and mixed Li--F--O species. Concentrated \ce{LiTFSI} electrolytes maintain multiple high-intensity fluorine layers, consistent with an early inorganic-rich interfacial layer, while \ce{LiPF6} displays a weaker and more diffuse fluorine distribution.

Figure~\ref{fig:Fig4}(E) shows the temporal evolution of the lithium crystalline structure during interaction with the different electrolyte formulations (\SI{1.5}{M}, \SI{2.5}{M}, and \SI{3.5}{M} \ce{LiTFSI}, and \SI{1}{M} \ce{LiPF6}), with corresponding cross-sectional snapshots provided in Figure~S9. A clear difference in interfacial amorphization is observed across the four systems. In the dilute and conventional regimes (\SI{1.5}{M} \ce{LiTFSI}, \SI{2.5}{M} \ce{LiTFSI}, and \SI{1}{M} \ce{LiPF6}), the native body-centered cubic lattice progressively degrades as solvent penetration disrupts the metallic surface, leading to disordered near-surface regions. By contrast, the Li-metal/\SI{3.5}{M} \ce{LiTFSI} interface exhibits greater crystalline persistence on the simulated timescale. Rather than penetrating as deeply into the bulk over the simulated timescale, inorganic species nucleate in a more layered manner that better retains the underlying lattice. Retention of a robust BCC core over the simulated timescale, up to \SI{200}{ps}, characterizes this early-growth mode. In the lower-concentration systems, the nascent SEI remains more permeable to solvent molecules, thereby exacerbating subsurface amorphization. These results indicate that structural evolution depends strongly on salt concentration, with the highly concentrated electrolyte forming a denser SEI that correlates with reduced deep amorphization in the simulated interface.

Corroborating these structural insights, Figure~\ref{fig:Fig4}(F) presents the mechanical response of the interfacial regions through simulated stress--strain profiles. These calculations are intended as comparative descriptors of the nascent interfacial regions formed within the simulation protocol. The \SI{3.5}{M} \ce{LiTFSI} system sustains higher ultimate stress than the lower-concentration benchmarks. This relative strengthening is consistent with the structural trends discussed above, as the denser layered nascent interphase may provide a more load-bearing framework within the simulation protocol. In contrast, the lower-concentration systems show varying degrees of mechanical softening associated with interfacial amorphization. Within this framework, the modulus increases from \SI{0.40}{GPa} for the \SI{1}{M} \ce{LiPF6} system to \SI{0.54}{GPa} and \SI{0.64}{GPa} for the \SI{1.5}{M} and \SI{2.5}{M} \ce{LiTFSI} systems, respectively, and reaches \SI{0.71}{GPa} for \SI{3.5}{M} \ce{LiTFSI}. These values should be interpreted only as relative stiffness indicators under the imposed simulation protocol and not as direct experimental elastic moduli. Within the imposed simulation protocol, the monotonic increase in modulus reflects a progressively more mechanically robust interphase as electrolyte concentration increases. This trend is broadly consistent with experimental PeakForce QNM measurements, which report SEI elastic moduli spanning \SIrange{0.2}{4.5}{GPa} \cite{shin2015component}.

\begin{figure*}[!ht]
    \centering
    \includegraphics[width=0.8\textwidth]{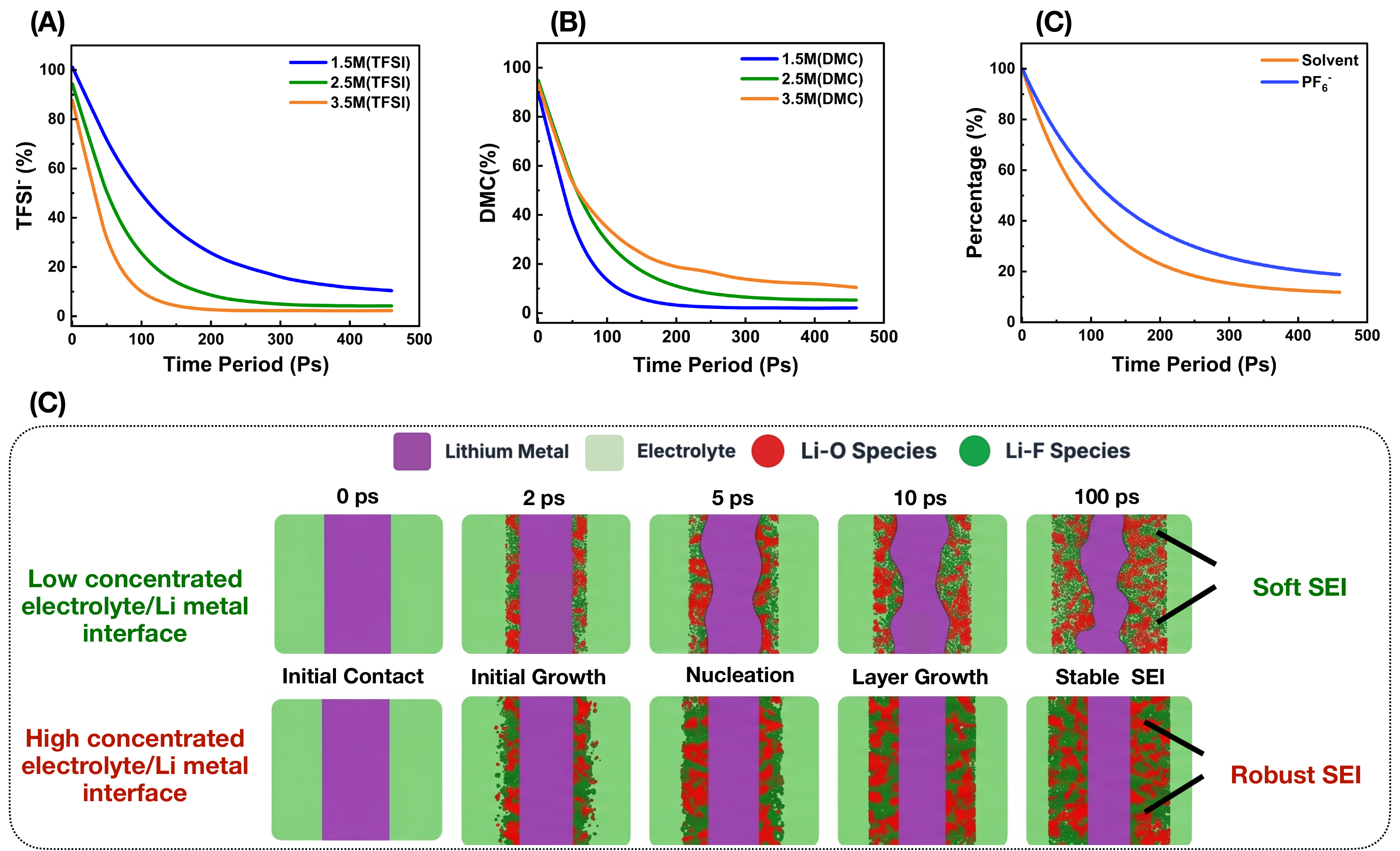}
    \caption{Species retention and schematic mechanism of SEI growth.
    (A) Fraction of intact \ce{TFSI-} remaining in the first solvation
    shell of \ce{Li+} as a function of simulation time for 1.5~M, 2.5~M,
    and 3.5~M \ce{LiTFSI}. (B) Corresponding retention of intact DMC
    solvent molecules in the \ce{Li+} solvation shell for the same three
    concentrations. (C) Retention of intact solvent and \ce{PF6-} species
    for the \SI{1}{M} \ce{LiPF6} reference system. (D) Schematic
    illustration summarizing the divergent mechanisms of SEI nucleation
    and growth observed across low- and high-concentration
    electrolyte/Li-metal interfaces.}
    \label{fig:Fig5}
\end{figure*}

Figure~\ref{fig:Fig5}(A--C) quantifies the loss of intact electrolyte
species from the \ce{Li+} solvation shell over time, complementing the
structural and compositional analyses above with a direct measure of
reactant consumption. For all three \ce{LiTFSI} concentrations
(Figure~\ref{fig:Fig5}A), the fraction of intact \ce{TFSI-} decays
rapidly within the first \SI{100}{ps} and then approaches a low
plateau by \SI{300}{ps}. The rate of this decay tracks concentration:
\SI{3.5}{M} \ce{LiTFSI} shows the fastest and most complete anion
consumption, falling to below 10\% intact \ce{TFSI-} within
\SI{100}{ps}, whereas \SI{1.5}{M} \ce{LiTFSI} decays more gradually and
retains a larger intact fraction (\(\sim\)10--20\%) even at
\SI{450}{ps}. This ordering is consistent with the anion-rich,
contact-ion-pair/aggregate-dominated solvation environment at high
concentration, which places more \ce{TFSI-} in direct proximity to the
Li surface and thus accelerates its reductive decomposition.

The DMC retention curves in Figure~\ref{fig:Fig5}(B) follow the same
qualitative trend, with solvent consumption also accelerating with
increasing \ce{LiTFSI} concentration, though the \SI{3.5}{M} and
\SI{2.5}{M} curves are closer to one another than the corresponding
\ce{TFSI-} curves, and both decay faster than the \SI{1.5}{M} case.
The similarity between the anion and solvent decay trends indicates
that, once the interphase begins to nucleate, both anion- and
solvent-derived pathways contribute to early consumption, with the
overall rate set primarily by how closely \ce{Li+} is coordinated by
reactive species in the bulk solvation shell.

For the \SI{1}{M} \ce{LiPF6} reference system (Figure~\ref{fig:Fig5}C),
both the solvent and \ce{PF6-} populations decay more slowly than any
of the \ce{LiTFSI} systems, with roughly 20\% of the solvent and
\(\sim\)20\% of \ce{PF6-} still intact at \SI{450}{ps}. This slower,
more gradual consumption is consistent with the weaker Li--F/Li--O
coordination and more diffuse fluorine and oxygen density profiles
observed for \ce{LiPF6} in Figure~\ref{fig:Fig4}, and reinforces the
picture of a less reactive, more slowly evolving nascent interphase
relative to concentrated \ce{LiTFSI}.

These trends are summarized schematically in Figure~\ref{fig:Fig5}(D).
The simulations indicate that highly concentrated \ce{LiTFSI} promotes
incipient formation of a denser early interphase within the first
\SIrange{2}{100}{ps}, consistent with the faster anion and solvent
consumption in panels A and B. This mechanistic picture aligns
qualitatively with reported electrochemical improvements in
concentrated electrolytes \cite{li2023concentrated}, although
long-term behavior is beyond the present timescale. In contrast,
\SI{1.5}{M} \ce{LiTFSI} and \SI{1}{M} \ce{LiPF6} produce more permeable
and mechanically weaker interphases within the simulation timescale,
consistent with their slower species consumption, which may influence
interfacial stability during extended cycling. The transition from
dense Li--O/F-rich nascent interphases in concentrated \ce{LiTFSI} to
more diffuse interphases in \ce{LiPF6} is qualitatively consistent with
reported XPS and TOF-SIMS characterizations
\cite{wang2018review,sayah2022super,xiao2025advances}. Because the
simulations capture only the earliest nucleation regime on
sub-nanosecond timescales, they describe incipient interphase
formation rather than fully matured SEI structure.\\

\section{Conclusion}
Bulk solvation architecture emerges as a predictive descriptor for early SEI nucleation timing. Increasing \ce{LiTFSI} concentration shifts the \ce{Li+} solvation shell toward contact ion pairs and aggregates, and this restructuring accelerates Li--F/Li--O coordination, drives faster conversion to inorganic LiF/Li$_2$O-rich products, and speeds consumption of intact \ce{TFSI-} and DMC, together yielding a denser, mechanically more robust nascent interphase within the first \SI{100}{ps}. More dilute \ce{LiTFSI} and \SI{1}{M} \ce{LiPF6} instead produce more diffuse, organic-rich, and softer interphases over the same window. Because these simulations resolve only the sub-nanosecond nucleation regime, longer trajectories and experimental validation are needed to connect this early behavior to long-term cycling stability; nonetheless, the results support solvation-guided electrolyte design as a practical handle for engineering more robust SEIs in lithium metal batteries.

\textbf{CRediT Authorship Contribution Statement.}
Syed Mustafa Shah and Mohammed Lemaalem: Conceptualization, Visualization, Software, Data Curation, Formal Analysis, Methodology, Writing -- Original Draft, Writing -- Review \& Editing. Anh T. Ngo: Supervision, Conceptualization, Data Curation, Formal Analysis, Investigation, Project Administration, Writing -- Review \& Editing.

\textbf{Author Contributions.}
This work was conceived by Anh T. Ngo. Syed Mustafa Shah and Mohammed Lemaalem carried out the calculations. All authors contributed to discussion of the results and revision of the manuscript.

\textbf{Declaration of Competing Interest.}
The authors declare no known competing financial interests or personal relationships that could have appeared to influence the work reported in this paper.

\textbf{Acknowledgements.}
The work reported in this paper was supported by the U.S. Department of Energy through the Battery Materials Research program. Argonne National Laboratory is operated for the U.S. Department of Energy by UChicago Argonne, LLC, under contract number DE-AC02-06CH11357. Computing resources were provided by the Laboratory Computing Resource Center at ANL.

\textbf{Supporting Information Available.}
Details on AIMD simulation parameters and cell configurations; the mathematical framework of the Deep Potential descriptor and training protocols; DP validations, including force/energy parity plots (Figures~S1--S6); full MLMD and CMD simulation methodologies; definitions of structural properties (RDF and coordination number) and transport properties (conductivity and viscosity); mean squared displacements of \ce{LiTFSI} and \ce{LiPF6} electrolytes at varying salt concentrations, along with corresponding MLMD simulation snapshots (Figure~S7); and additional structural analysis of early-stage SEI nucleation (Figures~S8 and S9).

\textbf{Data Availability.}
Data generated or analyzed during this study are available from the corresponding author upon reasonable request.

\bibliography{acs}

@article{li2023concentrated,
  title   = {A concentrated electrolyte of \ce{LiTFSI} and dimethyl carbonate for high-voltage 
  \ce{Li} batteries},
  author  = {Li, Yueqing and Wang, Zichen and Lin, Wentao and Wei, Bixia and Chen, Dengjie},
  journal = {ACS Applied Energy Materials},
  volume  = {6},
  number  = {18},
  pages   = {9337--9346},
  year    = {2023},
  publisher = {ACS Publications}
}

@article{ebadi2016electrolyte,
  title={Electrolyte decomposition on Li-metal surfaces from first-principles theory},
  author={Ebadi, Mahsa and Brandell, Daniel and Araujo, C Moyses},
  journal={The Journal of Chemical Physics},
  volume={145},
  number={20},
  year={2016},
  publisher={AIP Publishing}
}

@article{huang2024crucial,
  title   = {Crucial Roles of Ethyl Methyl Carbonate in {Lithium-Ion} and Dual-Ion Batteries: A Review},
  author  = {Huang, Yuhao and Luo, Yu and Wang, Binli and Wang, Hongyu and Zhang, Lei},
  journal = {Langmuir},
  volume  = {40},
  number  = {22},
  pages   = {11353--11370},
  year    = {2024},
  publisher = {ACS Publications}
}

@article{kuzmina2024structure,
  title   = {Structure and Physicochemical Properties of Solutions of Lithium Polysulfides in Tetrasolvate of Lithium Perchlorate with Sulfolane Molecular Dynamics Modeling},
  author  = {Kuzmina, Elena V and Yusupova, Alfia R and Karaseva, Elena V and Chen, Xiang and Zhang, Qiang and Kolosnitsyn, Vladimir S},
  journal = {The Journal of Physical Chemistry B},
  volume  = {128},
  number  = {32},
  pages   = {7833--7847},
  year    = {2024},
  publisher = {ACS Publications}
}

@article{jorge2024theoretically,
  title   = {Theoretically grounded approaches to account for polarization effects in fixed-charge force fields},
  author  = {Jorge, Miguel},
  journal = {The Journal of Chemical Physics},
  volume  = {161},
  number  = {18},
  year    = {2024},
  publisher = {AIP Publishing}
}

@article{seiferth2023limitations,
  title   = {Limitations of non-polarizable force fields in describing anion binding poses in non-polar synthetic hosts},
  author  = {Seiferth, David and Tucker, Stephen J and Biggin, Philip C},
  journal = {Physical Chemistry Chemical Physics},
  volume  = {25},
  number  = {26},
  pages   = {17596--17608},
  year    = {2023},
  publisher = {Royal Society of Chemistry}
}

@article{tao2025deep,
  title   = {Deep neural network enhanced mesoscopic thermodynamic model for unlocking the electrode/electrolyte interface},
  author  = {Tao, Haolan and Wang, Sijie and Liu, Honglai and Lian, Cheng},
  journal = {Angewandte Chemie International Edition},
  volume  = {64},
  number  = {6},
  pages   = {e202418447},
  year    = {2025},
  publisher = {Wiley Online Library}
}

@article{yao2022applying,
  title   = {Applying classical, ab initio, and machine-learning molecular dynamics simulations to the liquid electrolyte for rechargeable batteries},
  author  = {Yao, Nan and Chen, Xiang and Fu, Zhong-Heng and Zhang, Qiang},
  journal = {Chemical Reviews},
  volume  = {122},
  number  = {12},
  pages   = {10970--11021},
  year    = {2022},
  publisher = {ACS Publications}
}

@article{fan2024deep,
  title   = {Deep-learning-assisted insights into molecular transport in heterogeneous electrolyte films on electrodes},
  author  = {Fan, Linhao and Zuo, Ruiwang and Zhou, Yumeng and Ran, Aoxin and Li, Xing and Du, Qing and Jiao, Kui},
  journal = {Cell Reports Physical Science},
  volume  = {5},
  number  = {9},
  year    = {2024},
  publisher = {Elsevier}
}

@article{lai2025machine,
  title={Machine-learning-accelerated mechanistic exploration of interface modification in lithium metal anode},
  author={Lai, Genming and Zhang, Ruiqi and Fang, Chi and Zhao, Juntao and Chen, Taowen and Zuo, Yunxing and Xu, Bo and Zheng, Jiaxin},
  journal={npj Computational Materials},
  volume={11},
  number={1},
  pages={245},
  year={2025},
  publisher={Nature Publishing Group UK London}
}

@article{zeng2023deepmd,
  title   = {{DeePMD}-kit v2: A software package for deep potential models},
author = {Zeng, Jinzhe and Zhang, Duo and Lu, Denghui and Mo, Pinghui and Li, Zeyu and Chen, Yixiao and Rynik, Marián and Huang, Li’ang and Li, Ziyao and Shi, Shaochen and Wang, Yingze and Ye, Haotian and Tuo, Ping and Yang, Jiabin and Ding, Ye and Li, Yifan and Tisi, Davide and Zeng, Qiyu and Bao, Han and Xia, Yu and Huang, Jiameng and Muraoka, Koki and Wang, Yibo and Chang, Junhan and Yuan, Fengbo and Bore, Sigbjørn Løland and Cai, Chun and Lin, Yinnian and Wang, Bo and Xu, Jiayan and Zhu, Jia-Xin and Luo, Chenxing and Zhang, Yuzhi and Goodall, Rhys E. A. and Liang, Wenshuo and Singh, Anurag Kumar and Yao, Sikai and Zhang, Jingchao and Wentzcovitch, Renata and Han, Jiequn and Liu, Jie and Jia, Weile and York, Darrin M. and E, Weinan and Car, Roberto and Zhang, Linfeng and Wang, Han},
  journal = {The Journal of Chemical Physics},
  volume  = {159},
  number  = {5},
  year    = {2023},
  publisher = {AIP Publishing}
}

@article{zhang2018deep,
  title   = {End-to-end symmetry preserving inter-atomic potential energy model for finite and extended systems},
author = {Zhang, Linfeng and Han, Jiequn and Wang, Han and Saidi, Wissam and Car, Roberto and E, Weinan},
  journal = {Advances in Neural Information Processing Systems},
  volume  = {31},
  year    = {2018}
}

@article{xiao2025advances,
  title   = {Advances in anion chemistry in the electrolyte design for better lithium batteries},
  author  = {Xiao, Hecong and Li, Xiang and Fu, Yongzhu},
  journal = {Nano-Micro Letters},
  volume  = {17},
  number  = {1},
  pages   = {149},
  year    = {2025},
  publisher = {Springer}
}

@article{sayah2022super,
  title   = {How do super concentrated electrolytes push the Li-ion batteries and supercapacitors beyond their thermodynamic and electrochemical limits?},
  author  = {Sayah, Simon and Ghosh, Arunabh and Baazizi, Mariam and Amine, Rachid and Dahbi, Mouad and Amine, Youssef and Ghamouss, Fouad and Amine, Khalil},
  journal = {Nano Energy},
  volume  = {98},
  pages   = {107336},
  year    = {2022},
  publisher = {Elsevier}
}

@article{wang2018review,
  title   = {Review on modeling of the anode solid electrolyte interphase {(SEI)} for lithium-ion batteries},
  author  = {Wang, Aiping and Kadam, Sanket and Li, Hong and Shi, Siqi and Qi, Yue},
  journal = {npj Computational Materials},
  volume  = {4},
  number  = {1},
  pages   = {15},
  year    = {2018},
  publisher = {Nature Publishing Group UK London}
}

@article{yang2021robust,
  title   = {Robust solid/electrolyte interphase {(SEI)} formation on Si anodes using glyme-based electrolytes},
  author = {Yang, Guang and Frisco, Sarah and Tao, Runming and Philip, Nathan and Bennett, Tyler H. and Stetson, Caleb and Zhang, Ji-Guang and Han, Sang-Don and Teeter, Glenn and Harvey, Steven P. and Zhang, Yunya and Veith, Gabriel M. and Nanda, Jagjit},


  journal = {ACS Energy Letters},
  volume  = {6},
  number  = {5},
  pages   = {1684--1693},
  year    = {2021},
  publisher = {ACS Publications}
}

@article{alvarado2019bisalt,
  title   = {Bisalt ether electrolytes: a pathway towards lithium metal batteries with Ni-rich cathodes},
author ="Alvarado, Judith and Schroeder, Marshall A. and Pollard, Travis P. and Wang, Xuefeng and Lee, Jungwoo Z. and Zhang, Minghao and Wynn, Thomas and Ding, Michael and Borodin, Oleg and Meng, Ying Shirley and Xu, Kang",
  journal = {Energy \& Environmental Science},
  volume  = {12},
  number  = {2},
  pages   = {780--794},
  year    = {2019},
  publisher = {Royal Society of Chemistry}
}

@article{shadike2022engineering,
  title   = {Engineering and characterization of interphases for lithium metal anodes},
  author  = {Shadike, Zulipiya and Tan, Sha and Lin, Ruoqian and Cao, Xia and Hu, Enyuan and Yang, Xiao-Qing},
  journal = {Chemical Science},
  volume  = {13},
  number  = {6},
  pages   = {1547--1568},
  year    = {2022},
  publisher = {Royal Society of Chemistry}
}

@article{gao2025lif,
  title   = {LiF Artifacts in XPS Analysis of the {SEI} for Lithium Metal Batteries},
  author  = {Gao, Aosong and Lai, Hao and Duan, Mingqiu and Chen, Si and Huang, Wenyu and Yang, Muzi and Gong, Li and Chen, Jian and Xie, Fangyan and Meng, Hui},
  journal = {ACS Applied Materials \& Interfaces},
  volume  = {17},
  number  = {5},
  pages   = {8513--8525},
  year    = {2025},
  publisher = {ACS Publications}
}

@article{kuai2023inorganic,
  title   = {Inorganic solid electrolyte interphase engineering rationales inspired by hexafluorophosphate decomposition mechanisms},
  author  = {Kuai, Dacheng and Balbuena, Perla B.},
  journal = {The Journal of Physical Chemistry C},
  volume  = {127},
  number  = {4},
  pages   = {1744--1751},
  year    = {2023},
  publisher = {ACS Publications}
}

@article{Zhong2025MLSEI,
  title        = {Machine-Learning-Guided Insights into Solid-Electrolyte Interphase Conductivity: Are Amorphous Lithium Fluorophosphates the Key?},
  author       = {Zhong, Peichen and Persson, Kristin A.},
  journal      = {ACS Energy Letters},
  volume = {11},
number = {1},
pages = {806-812},
year = {2026}
}

@article{zhou2025tailored,
  title={Tailored Engineering on the Interface Between Lithium Metal Anode and Solid-State Electrolytes},
  author={Zhou, Qi and Xiong, Xiaosong and Peng, Jun and Wu, Wenzhuo and Fan, Weijia and Yang, Haoyuan and Wang, Tao and Ma, Yuan and Wang, Faxing and Wu, Yuping},
  journal={Energy \& Environmental Materials},
  volume={8},
  number={1},
  pages={e12831},
  year={2025},
  publisher={Wiley Online Library}
}

@article{lindsey2026situ,
  title={In situ and operando microscopy studies on lithium metal anodes: a review},
  author={Lindsey, Ian and Mondl, Cameron and Meng, Xiangbo},
  journal={Energy Advances},
  year  ="2026",
volume  ="5",
issue  ="1",
pages  ="7-42",
publisher  ="RSC",
doi  ="10.1039/D5YA00240K"
}

@article{wu2025formation,
  title={Formation mechanisms of solid electrolyte interphase and its influence on lithium battery performance},
  author={Wu, Yinglei and Ge, Guangfu and Wang, Sirui and Xiong, Liping and He, Zhongyi},
  journal={Materials Today Energy},
  pages={102124},
  year={2025},
  publisher={Elsevier}
}

@article{zhang2025recent,
  title={Recent progress in constructing fluorinated solid-electrolyte interphases for stable lithium metal anodes},
  author={Zhang, Di and Lv, Pengfei and Qin, Wei and He, Xin and He, Yuanhua},
  journal={International Journal of Minerals, Metallurgy and Materials},
  volume={32},
  number={2},
  pages={270--291},
  year={2025},
  publisher={Springer}
}

@article{choi2025strategic,
title={Strategic Surface Engineering of Lithium Metal Anodes: Simultaneous Native Layer Elimination and Protective Layer Formation via Gas--Solid Reaction},
author={Choi, Siwon and Chae, Seongwook and Kim, Taemin and Shin, Hyeonsol and Bae, Jin-Gyu and Lee, Seung Geol and Lee, Ji Hoon and Lee, Hyeon Jeong},
journal={ACS nano},
volume={19},
number={16},
pages={16119--16132},

  year={2025},

  publisher={ACS Publications}

}

@article{cheng2025spontaneous,

  title={Spontaneous Formation of Robust Hybrid Organic/Inorganic Interface for Advancing Lithium Metal Batteries},

  author = {Cheng, Hanwen and Shen, ChunLi and Cui, Lianmeng and Li, Jinghao and Ren, Long and Dong, Chenxu and Hu, Wenxi and Zhang, Wenwei and Zhou, Min and Xiong, Yan and Liao, Xiaobin and Zhao, Yan},

  journal={ACS Applied Materials \& Interfaces},

  volume={17},

  number={15},

  pages={22677--22686},

  year={2025},

  publisher={ACS Publications}

}

@article{perez2024sei,

  title={{SEI} formation and lithium-ion electrodeposition dynamics in lithium metal batteries via first-principles kinetic monte carlo modeling},

  author={Perez-Beltran, Saul and Kuai, Dacheng and Balbuena, Perla B},

  journal={ACS Energy Letters},

  volume={9},

  number={11},

  pages={5268--5278},

  year={2024},

  publisher={ACS Publications}

}

@article{beltran2022sei,

  title={{SEI} formation mechanisms and \ce{Li+} dissolution in lithium metal anodes: impact of the electrolyte composition and the electrolyte-to-anode ratio},

  author={Beltran, Saul Perez and Balbuena, Perla B},

  journal={Journal of Power Sources},

  volume={551},

  pages={232203},

  year={2022},

  publisher={Elsevier}

}

@article{niu2025large,
  title={A large-scale on-the-fly machine learning molecular dynamics simulation to explore lithium metal battery interfaces},
  author={Niu, Yi-Lin and Chen, Xiang and Zhang, Tian-Chen and Gao, Yu-Chen and Chen, Yao-Peng and Yao, Nan and Fu, Zhong-Heng and Zhang, Qiang},
  journal={Journal of Energy Chemistry},
  year={2025},
  publisher={Elsevier}
}

@article{Wang2025_RecentTrends,
  title={Recent advances in the stabilization of lithium metal anodes: From solvation chemistry to interfacial engineering},
  author={Wang, H. and Lee, S. and Smith, J.},
  journal={Journal of Energy Chemistry},
  volume={102},
  pages={45--68},
  year={2025},
  publisher={Elsevier}
}

@article{lee2026recent,
  title={Recent Advances in Lithium Metal Anodes with Liquid Electrolytes: Interfacial Interaction-Driven Assembly for Dendrite Suppression and Long-Term Stability},
  author={Lee, Chanseok and Kim, Boyeon and Hwang, Hyunjun and Moon, Jun Hyuk and Chung, Yoon Jang and Kim, Hyeong Jun and Ko, Yongmin and Cho, Jinhan},
  journal={Small},
  pages={e13356},
  year={2026},
  publisher={Wiley Online Library}
}

@article{vishweswariah2025evaluation,
  title={Evaluation and Characterization of SEI Composition in Lithium Metal and Anode-Free Lithium Batteries},
  author={Vishweswariah, Karthik and Ningappa, Ningaraju Gejjiganahalli and Bouguern, Mohamed Djihad and Kumar MR, Anil and Armand, Michel B and Zaghib, Karim},
  journal={Advanced Energy Materials},
  volume={15},
  number={39},
  pages={2501883},
  year={2025},
  publisher={Wiley Online Library}
}

@article{xing2025directing,
  title={Directing selective solvent presentations at electrochemical interfaces to enable initially anode-free sodium metal batteries},
  author={Xing, Qianli and Lee, Jung Min and Yang, Ziqi and Van Lehn, Reid C and Liu, Fang},
  journal={Nature communications},
  volume={16},
  number={1},
  pages={8265},
  year={2025},
  publisher={Nature Publishing Group UK London}
}

@article{li2026non,
  title={Non-fluorinated electrolyte for high-voltage anode-free sodium metal battery},
  author={Li, Ai-Min and Pollard, Travis P and Wang, Zeyi and Zhang, Nan and Omenya, Fred and Tan, Sha and Hu, Enyuan and Yang, Xiao-Qing and Li, Xiaolin and Borodin, Oleg and others},
  journal={Nature Sustainability},
  volume={9},
  number={2},
  pages={306--316},
  year={2026},
  publisher={Nature Publishing Group UK London}
}

@article{king2025machine,
  title={Machine learning of charges and long-range interactions from energies and forces},
  author={King, Daniel S and Kim, Dongjin and Zhong, Peichen and Cheng, Bingqing},
  journal={Nature Communications},
  volume={16},
  number={1},
  pages={8763},
  year={2025},
  publisher={Nature Publishing Group UK London}
}

@article{shin2015component,
  title={Component-/structure-dependent elasticity of solid electrolyte interphase layer in Li-ion batteries: Experimental and computational studies},
  author={Shin, Hosop and Park, Jonghyun and Han, Sangwoo and Sastry, Ann Marie and Lu, Wei},
  journal={Journal of Power Sources},
  volume={277},
  pages={169--179},
  year={2015},
  publisher={Elsevier}
}
\newpage
\includepdf[pages=-]{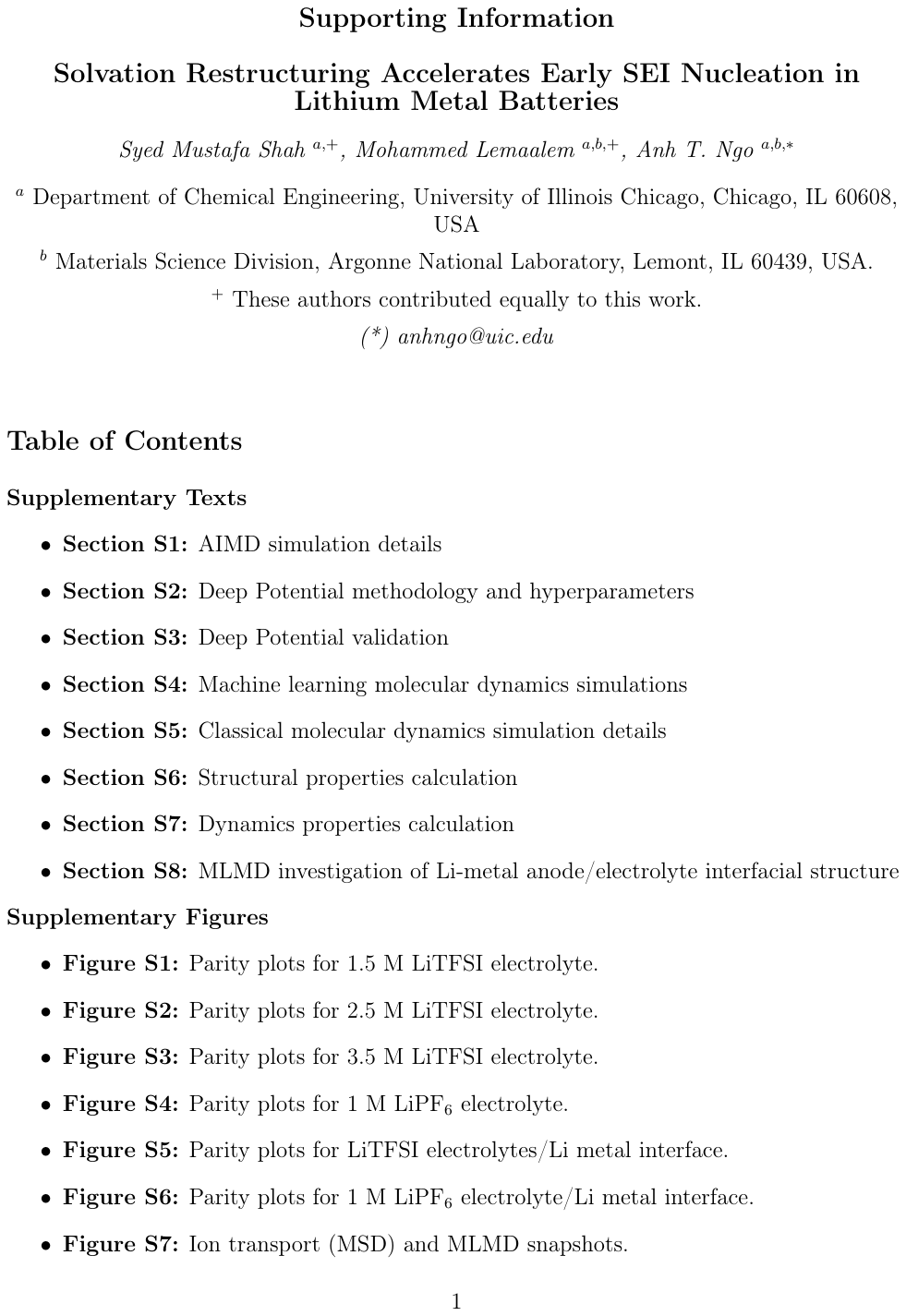} 

\end{document}